\documentclass[prd,twocolumn,aps,showpacs,preprintnumbers,amsmath,amssymb]{revtex4}
\usepackage{graphicx}
\usepackage{dcolumn}
\usepackage{bm}
\usepackage{psfrag}

\begin{document}
\newcommand{\be}{\begin{equation}}\newcommand{\ee}{\end{equation}}
\newcommand{\bea}{\begin{eqnarray}}\newcommand{\eea}{\end{eqnarray}}
\newcommand{\bc}{\begin{center}}\newcommand{\ec}{\end{center}}
\def\no{\nonumber}
\def\eq#1{Eq. (\ref{#1})}\def\eqeq#1#2{Eqs. (\ref{#1}) and  (\ref{#2})}
\def\lsim{\raise0.3ex\hbox{$\;<$\kern-0.75em\raise-1.1ex\hbox{$\sim\;$}}}
\def\gsim{\raise0.3ex\hbox{$\;>$\kern-0.75em\raise-1.1ex\hbox{$\sim\;$}}}
\def\slash#1{\ooalign{\hfil/\hfil\crcr$#1$}}
\def\eff{\mbox{\tiny{eff}}}
\def\order#1{{\mathcal{O}}(#1)}
\def\pppm{B^0\to\pi^+\pi^-}
\def\pzpz{B^0\to\pi^0\pi^0}
\def\pppz{B^0\to\pi^+\pi^0}
\preprint{ }
\title{${\mathcal A}_{\rm CP}$ Puzzle~: Possible Evidence for Large
  Strong Phase in\\ $B\to K\pi$ Color-Suppressed Tree Amplitude}
\author{
T. N. Pham}
\affiliation{
 Centre de Physique Th\'{e}orique, CNRS \\ 
Ecole Polytechnique, 91128 Palaiseau, Cedex, France }
\date{\today}
\begin{abstract}
In QCD Factorization(QCDF),  the suppression of the color-suppressed 
tree amplitude relative to the color-allowed one in $B\to K\pi$ decay 
implies a direct CP asymmetry in $B^{-}\to K^{-}\pi^{0}$
to be of the same sign and comparable in magnitude to that in 
${\bar B}^{0}\to K^{+}\pi^{-}$, in contradiction with experiment. This is
the $A_{\rm CP}$ $B\to K\pi$ puzzle. One of the current proposal to solve 
this puzzle is the existence of a large color-suppressed amplitude 
with  large strong phase which implies also a large negative 
${\bar B}^{0}\to {\bar K}^{0}\pi^{0}$ CP asymmetry. In this paper, 
by  an essentially model-independent calculation, we show clearly that
the large negative direct CP asymmetry in ${\bar B}^{0}\to {\bar K}^{0}\pi^{0}$
implies a large $C/T$, the ratio of the color-suppressed to the
color-allowed tree amplitude and a large negative strong phase 
for $C$. By adding to the  QCDF amplitude  an additional color-suppressed
term to generate a large $C/T$ and a large strong phase for $C$ and 
an additional penguin-like contribution,  we obtain  branching ratios 
for all  $B\to K\pi$ modes and CP asymmetry for ${\bar B}^{0}\to
K^{+}\pi^{-}$ and $B^{-}\to K^{-}\pi^{0}$ in agreement  with experiment,
and a large and negative CP asymmetry in 
${\bar B}^{0}\to {\bar K}^{0}\pi^{0}$ which could be checked with 
more precise measurements.

\end{abstract}
\pacs{13.25.Hw, 12.38.Bx}
\maketitle
\section{INTRODUCTION}
In the penguin-dominated $B\to K\pi$ decays, the color-suppressed tree
contribution($C$) is suppressed relative to the color-allowed
tree contribution($T$) because of the small Wilson coefficient $a_{2}$
relative to $a_{1}$. One would then expect the direct CP
asymmetries in   $B^{-}\to K^{-}\pi^{0}$ and ${\bar B}^{0}\to K^{-}\pi^{+}$
to be essentially given by the color-allowed tree and strong penguin 
interference terms($TP$). The CP asymmetry(${\mathcal A}_{\rm CP}$) 
in $B^{-}\to K^{-}\pi^{0}$ would be of the same sign and comparable 
in magnitude to that in 
${\bar B}^{0}\to K^{-}\pi^{+}$. The current measurements\cite{HFAG}, 
though with large errors, seem to indicate a positive CP asymmetry 
for $B^{-}\to K^{-}\pi^{0}$, in opposite sign to the negative
 ${\bar B}^{0}\to K^{-}\pi^{+}$  CP  asymmetry measured 
with greater accuracy. This is the $A_{\rm CP}$
puzzle\cite{Li1,Kim,Li2,Fleischer1,Rosner1,Fleischer,Kim2,Gronau1,Ciuchini1,Chiang,Chiang2}. 

  To reverse the sign of the predicted 
$B^{-}\to K^{-}\pi^{0}$ CP asymmetry, one would need a large  
color-suppressed tree terms, i.e a large $C/T$ ratio, and also a large
strong phase for $C$, as will be shown in the following. Since
the color-suppressed tree-penguin  interference term 
in $B^{-}\to K^{-}\pi^{0}$ is opposite in sign to that 
in ${\bar B}^{0}\to {\bar K}^{0}\pi^{0}$, the
 ${\bar B}^{0}\to {\bar K}^{0}\pi^{0}$ CP asymmetry would become
large and negative. If the positive asymmetry for $B^{-}\to K^{-}\pi^{0}$
and a large negative ${\bar B}^{0}\to {\bar K}^{0}\pi^{0}$ CP asymmetry are
confirmed by new measurements, this would be a clear evidence for 
the enhanced color-suppressed tree contribution to CP asymmetries in 
$B\to K\pi$ decays. Apart from the possibility
of new physics\cite{Fleischer,Kim2,Khalil} to solve the $A_{\rm CP}$
$B\to K\pi$ puzzle, recent calculations in the standard model(SM), as
done  in perturbative QCD (pQCD)\cite{Li1,Li2}, in QCD Factorization(QCDF) 
with large hard scattering
corrections\cite{Cheng2} seem to obtain a large color-suppressed
enhancement in  $B\to \pi\pi$ and $B\to K\pi$ decays. The  calculation 
in \cite{Kim2} also shows that  
the color-suppressed tree contribution has to be large to solve
the $B\to K\pi$ $A_{\rm CP}$ puzzle within the standard model.
 Various phenomenological
analyses\cite{Rosner2} using $SU(3)$ symmetry obtain also a large $C/T$
ratio. Final state interaction (FSI) rescattering term  with
a large absorptive part, like the charmed meson rescattering
charming penguin contribution\cite{Isola,Ciuchini,Santorelli,Santorelli1,Fazio,Fazio2,Cheng}, could also produce a large $C/T$ with a strong 
phase\cite{Kim,Cheng,Fajfer,Kou}, for example, through the 
CKM-suppressed, color-allowed
tree rescattering $B\to K^{*}\rho \to K\pi$ process, which produces
a tree-penguin interference term responsible for CP asymmetry, similar to the 
process $B\to \rho\rho \to \pi\pi$ in $B\to \pi\pi$ decays. Before going
further in analyzing these possibilities, one would like to have a
model-independent calculation to show that, apart from the possibility  of
new physics,  the solution to the 
$A_{\rm CP}$ puzzle is  an enhanced color-suppressed contribution to
CP asymmetry in $B\to K\pi$ decays. In the next section  we will show in an
essentially model-independent calculation that the large negative
${\bar B}^{0}\to {\bar K}^{0}\pi^{0}$ CP asymmetry requires a large
ratio $C/T$ with $C$ mainly absorptive. We then show  that with this 
additional contribution to the color-suppressed tree term and a penguin-like
additional term as given in \cite{Pham3} , QCDF could 
predict  all the branching ratios and CP asymmetries for $B\to K\pi$ decays 
consistent with experiment. 

\section{MODEL-INDEPENDENT DETERMINATION OF STRONG PHASES IN $B\to K\pi$}
  As our analysis is based on QCD Factorization,
for convenience, we reproduce here the QCDF $B\to K\pi$ decay amplitudes
given in \cite{Pham3}. We have\cite{Ali,QCDF1,QCDF2,Zhu3}~:
\bea
\kern -0.6cm&&A(B^{-}\to K^{-}\pi^{0}) = -i\frac{G_{F}}{2}f_{K}F^{B\pi}_{0}(m^{2}_{K})
(m_{B}^{2}-m_{\pi}^{2})\nonumber \\
\kern -0.6cm&&\left(V_{ub}V^{*}_{us}a_{1}
+(V_{ub}V^{*}_{us}+V_{cb}V^{*}_{cs})[a_{4} + a_{10} + (a_{6}+a_{8})r_{\chi}]\right)\nonumber \\ 
\kern -0.6cm&& -i\frac{G_{F}}{2}f_{\pi}F^{BK}_{0}(m^{2}_{\pi})(m_{B}^{2}-m_{K}^{2})\nonumber \\ 
\kern -0.6cm
&&\times\left(V_{ub}V^{*}_{us}a_{2}+(V_{ub}V^{*}_{us}+V_{cb}V^{*}_{cs})\times
  \frac{3}{2}(a_{9}-a_{7})\right)\nonumber \\ 
\kern -0.6cm&&-i\frac{G_{F}}{2}f_{B}f_{K}f_{\pi}\nonumber \\ 
\kern -0.6cm&&\times\left[V_{ub}V^{*}_{us}b_{2} 
+ (V_{ub}V^{*}_{us}+V_{cb}V^{*}_{cs})\times(b_{3} + b_{3}^{ew})\right]
\label{K1p0} \\
\kern -0.6cm&&A(B^{-}\to {\bar K}^{0}\pi^{-}) =-i\frac{G_{F}}{\sqrt{2}}f_{K}F^{B\pi}_{0}(m^{2}_{K})(m_{B}^{2}-m_{\pi}^{2})\nonumber \\
\kern -0.6cm&&+(V_{ub}V^{*}_{us}+V_{cb}V^{*}_{cs})\left[a_{4} - \frac{1}{2}a_{10} + (a_{6}-\frac{1}{2}a_{8})r_{\chi}\right]\nonumber \\
\kern -0.6cm&&-i\frac{G_{F}}{\sqrt{2}}f_{B}f_{K}f_{\pi}\nonumber \\ 
\kern -0.6cm&&\times\left[V_{ub}V^{*}_{us}b_{2} 
+ (V_{ub}V^{*}_{us}+V_{cb}V^{*}_{cs})\times(b_{3} + b_{3}^{ew})\right]
\label{K0p1}
\eea
and for ${\bar B}^{0}$ :
\bea
\kern -0.6cm&&A({\bar B}^{0}\to K^{-}\pi^{+}) = -i\frac{G_{F}}{\sqrt{2}}f_{K}F^{B\pi}_{0}(m^{2}_{K})(m_{B}^{2}-m_{\pi}^{2})\nonumber \\
\kern -0.6cm&&\biggl(\kern -0.1cm V_{ub}V^{*}_{us}a_{1}+\kern -0.1cm (V_{ub}V^{*}_{us}+\kern -0.1cm
  V_{cb}V^{*}_{cs})[a_{4} +\kern -0.1cm a_{10} + \kern
-0.1cm(a_{6}+a_{8})r_{\chi}]\kern -0.1cm\biggr)\nonumber\\
\kern -0.6cm&&-i\frac{G_{F}}{\sqrt{2}}f_{B}f_{K}f_{\pi}\left[
 (V_{ub}V^{*}_{us}+V_{cb}V^{*}_{cs})\times(b_{3} - \frac{b_{3}^{ew}}{2})\right]
\label{K1p2}\\
\kern -0.6cm&&A({\bar B}^{0}\to {\bar K}^{0}\pi^{0}) =i\frac{G_{F}}{2}f_{K}F^{B\pi}_{0}(m^{2}_{K})(m_{B}^{2}-m_{\pi}^{2})\nonumber \\
\kern -0.6cm&&\times(V_{ub}V^{*}_{us}+V_{cb}V^{*}_{cs})\left[a_{4} - \frac{1}{2}a_{10} + (a_{6}-\frac{1}{2}a_{8})r_{\chi}\right]\nonumber \\
\kern -0.6cm&& -i\frac{G_{F}}{2}f_{\pi}F^{BK}_{0}(m^{2}_{\pi})(m_{B}^{2}-m_{K}^{2})\nonumber \\ 
\kern -0.6cm&&\left(V_{ub}V^{*}_{us}a_{2}+(V_{ub}V^{*}_{us}+V_{cb}V^{*}_{cs})\times \frac{3}{2}(a_{9}-a_{7})\right)\nonumber \\ 
\kern -0.6cm&&+i\frac{G_{F}}{2}f_{B}f_{K}f_{\pi}\left[
 (V_{ub}V^{*}_{us}+V_{cb}V^{*}_{cs})\times(b_{3} - \frac{b_{3}^{ew}}{2})\right]
\label{K0p0}
\eea
where $r_{\chi}=\frac{2m_{K}^{2}}{(m_{b}-m_{d})(m_{d} + m_{s})}$
is the chirally-enhanced terms in the penguin $O_{6}$ matrix
element. The annihilation term $b_{i}$ are evaluated with the 
factor $f_{B}f_{M_{1}}f_{M_{2}}$ included and normalized relative 
to the factor $f_{K}F^{B\pi}_{0}(m_{B}^{2}-m_{\pi}^{2})$ in the 
factorisable terms. For the $B^{-}\to \pi^{-}\pi^{0} $ amplitude, we have:
\bea
\kern -0.6cm &&A(B^{-}\to \pi^{-}\pi^{0}) = -i\frac{G_{F}}{2}f_{\pi}F^{B\pi}_{0}(m^{2}_{\pi})
(m_{B}^{2}-m_{\pi}^{2})\nonumber \\
&&\biggl(V_{ub}V^{*}_{ud}(a_{1} + a_{2})
+(V_{ub}V^{*}_{ud} +V_{cb}V^{*}_{cd})\nonumber \\
\kern -0.6cm&&\times \frac{3}{2}(a_{9}-a_{7}+ a_{10} +a_{8}r_{\chi})\biggr)
\label{p2p1}
\eea

We see that the $B\to K\pi$ decay
amplitudes consist of a QCD penguin(P) $a_{4} + a_{6}r_{\chi} $,
a color-allowed tree(T) $a_{1}$, a color-suppressed  tree(C) $a_{2}$
, a color-allowed electroweak penguin(EW) $a_{9}-a_{7}$, a 
color-suppressed  electroweak penguin(EWC) $a_{10}+ a_{8}r_{\chi}$ terms. 
(There are also the penguin contribution given by  $a^{u}_{4} +
a^{u}_{6}r_{\chi} $ term not shown in the above expressions, for simplicity).
Because of the relative large Wilson coefficients, the QCD penguin, the  
color-allowed tree and the color-allowed electroweak contribution 
are the major contributions in $B\to K\pi$ decays. The  $B\to K\pi$
amplitude in Eqs.(\ref{K1p0})-(\ref{K0p0}) are 
then given as the
sum of the allowed-tree $T$, the color-suppressed tree $C$, the
color-allowed electroweak penguin $P_{\rm W}$ , the color-suppressed  
electroweak penguin,  $P_{\rm WC}$, 
tree-annihilation $A$ (the $b_{2}$ terms in Eq.(\ref{K1p0})-(\ref{K0p1}))  
the penguin-induced weak annihilation $P_{\rm A}$.
One can further simplify the  expressions, by grouping together
the penguin and penguin weak annihilation as an effective penguin
$P_{\rm eff}$ as usually done\cite{QCDF2}, furthermore, since
the CKM-suppressed, color-suppressed $b_{2}$ terms are much smaller than
the color-allowed tree term, we could also neglect $A$, and put the 
tree terms and the CKM-suppressed part of $P$ and $P_{A}$ into an
effective $T_{\rm eff}$  and $C_{\rm eff}$ . The $B\to K\pi$
amplitudes in terms of the effective penguin and tree amplitude are
then (putting $P_{\rm eff}=P$,  $T_{\rm eff}=T$, $C_{\rm eff}=C$), 
we have (in the notations of Ref.\cite{Cheng2}):
\bea
 A(B^-\to K^-\pi^0)  &=&
\frac{1}{\sqrt{2}}(P e^{i\delta_P}+ Te^{i\delta_T}e^{i\gamma}+
Ce^{i\delta_C}e^{-i\gamma} \nonumber \\
&+& P_{\rm W}+\frac{2}{3}P_{\rm WC}), \nonumber \\
A(B^-\to \bar K^0\pi^-) &=&  P e^{i\delta_P}-\frac{1}{3}P_{\rm WC},  \nonumber \\
 A(\bar B^0\to K^-\pi^+) &=&\kern -0.1cm P e^{i\delta_P}+T e^{i\delta_T}e^{-i\gamma}+\frac{2}{3}P_{\rm WC}, \nonumber \\
A(\bar B^0\to \bar K^0\pi^0)\kern -0.1cm &=& -\frac{1}{\sqrt{2}}(P e^{i\delta_P}- C e^{i\delta_C}e^{-i\gamma}\nonumber \\
&-& P_{\rm W}-\frac{1}{3}P_{\rm  WC}) .
\label{AKpi}
\eea
with the strong phase  $\delta_{P},\delta_{T},\delta_{C}$
for the penguin, color-allowed and color-suppressed tree, respectively and
the  weak phase $\gamma$ of the CKM matrix element
$V_{ub}$ is written explicitly in
the color-allowed $T$ and color-suppressed $C$ terms. In terms of the
relative strong phase $\delta_{PT}$, $\delta_{CT}$, and 
to take into account of the fact that the real part of the penguin
 amplitude $P$ is negative  in QCDF, we have
 $\delta_{P}= \delta_{PT}+ \pi + \delta_{T}$, and
$\delta_{C}= \delta_{CT} + \delta_{T}$. 

  Consider now the CP-averaged
$\Gamma_{\rm av} $and  CP-difference $\Gamma_{\rm as}$ for $B\to K\pi$ 
decay rates are then, with 
$\Gamma_{\rm av}= (\Gamma(B\to K\pi) + {\bar \Gamma}(B\to K\pi))/2$, 
$\Gamma_{\rm as}= (\Gamma(B\to K\pi) - {\bar \Gamma}(B\to K\pi))$ and
$\Gamma(B\to K\pi)$ and ${\bar \Gamma}(B\to K\pi)$ denotes the 
 decay rate for the corresponding charge-conjugate mode. We have
\bea
\kern -0.6cm &&\Gamma_{\rm av}(B^-\to K^-\pi^0)  =\frac{ P^2}{2}
- P\,T\cos(\delta_{PT})\,\cos(\gamma)\nonumber\\&& - P\,C\cos(\delta_{PT}-\delta_{CT})\,\cos(\gamma)+T\,C\cos(\delta_{CT})\nonumber\\
&&- P\,P_{W}\cos(\delta_{PT}+\delta_T)\,\cos(\gamma)
+ T\,P_{W}\cos(\delta_T)\,\cos(\gamma)
\nonumber\\&&+C\,P_{W}\cos(\delta_{CT}+\delta_T)\,\cos(\gamma)+\frac{T^2}{2}+ 
\frac{C^{2}}{2}
+ \frac{P_{W}^2}{2}  \label{br1} \\
\kern -0.6cm&& \Gamma_{\rm av}(B^-\to \bar K^0\pi^-) =  P^{2}  \label{br2} \\
\kern -0.6cm&&\Gamma_{\rm av}(\bar B^0\to K^-\pi^+) = P^2
-2\,P\,T\cos(\delta_{PT})\,\cos(\gamma) + \nonumber\\&& \kern 3cm+ T^2
 \label{br3} \\
\kern -0.6cm&&\Gamma_{\rm av }(\bar B^0\to \bar K^0\pi^0)= \frac{ P^2}{2}
+ P\,C\cos(\delta_{PT}-\delta_{CT})\,\cos(\gamma)\nonumber\\
&&\kern 1.0cm + P\,P_{W}\cos(\delta_{PT}+\delta_T)\,\cos(\gamma)\nonumber\\
&&\kern 1.0cm + C\,P_{W}\cos(\delta_{CT}+\delta_T)\,\cos(\gamma) +\frac{C^2}{2}+ \frac{P_{W}^2}{2}
\label{br4}
\eea
where $P$, $T$, $C$ and $P_{W}$ are positive and the negative real part 
of the penguin term has been taken into account in the phase 
$\delta_{P}= \pi + \delta_{PT}+\delta_{T}$ as mentioned above. Also, 
to simplify the analysis, we have neglected the color-suppressed electroweak
penguin $P_{WC}$ contribution which is smaller than the color-allowed 
elctroweak penguin $P_{W}$ by an order of magnitude as can be seen 
from the $a_{8}$ and $a_{10}$ terms in Eqs.(\ref{K1p0})-(\ref{K0p0}). For the
CP-difference decay rates, we obtain:
\bea
\kern -0.6cm &&\Gamma_{\rm as}(B^-\to K^-\pi^0)  =
2\,P\,T\sin(\delta_{PT})\,\sin(\gamma)\nonumber\\&&
+2\,P\,C\sin(\delta_{PT}-\delta_{CT})\,\sin(\gamma) +
2\,T\,P_{W}\sin(\delta_T)\,\sin(\gamma)\nonumber\\&&
\kern 1.0cm+ 2\,C\,P_{W}\sin(\delta_{CT}+\delta_T)\,\sin(\gamma)\ , 
\label{as1} \\
\kern -0.6cm&& \Gamma_{\rm as}(B^-\to \bar K^0\pi^-) = 0 \ , 
\label{as2} \\
\kern -0.6cm&&\Gamma_{\rm as}(\bar B^0\to K^-\pi^+)=
4\,P\,T\sin(\delta_{PT})\,\sin(\gamma)  \quad,  \label{as3} \\
\kern -0.6cm&&\Gamma_{\rm as }(\bar B^0\to \bar K^0\pi^0)=
- 2\,P\,C\sin(\delta_{PT}-\delta_{CT})\,\sin(\gamma)
\nonumber\\&&\kern 3.0cm + 2\,C\,P_{W}\sin(\delta_{CT}+\delta_T)\,\sin(\gamma)  
. \label{as4}
\eea
and the CP asymmetries are then given by:
\be
{\mathcal A}_{\rm CP}(B\to K\pi)= \frac{\Gamma_{\rm as}(B\to
  K\pi)}{2\Gamma_{\rm av}(B\to K\pi)}
\label{ACP}
\ee
As the $B\to K\pi$ branching ratios have been measured with an accuracy 
at the $10^{-6}$ level, it is possible to use the differences in the 
measured branching ratios and CP asymmetry to determine the 
relative $T/P$, $C/T$ and the strong phase $\delta_{PT}$, $\delta_{CT}$, 
as done for $B\to \pi\pi$ decays\cite{Kou,WW} in which the relative
strong phase $\delta_{PT}$ can be extracted from the measured
mixing-induced and direct CP asymmetry parameters $S_{\pi^{+}\pi^{-}}$ and  
$C_{\pi^{+}\pi^{-}}$ . For example, by neglecting  the $(P/T)^{2}$
term in $S_{\pi^{+}\pi^{-}}$, one would obtain:
 \be
\kern0.5cm \tan\delta_{PT} \approx - C_{\pi^+\pi^-}/ S_{\pi^+\pi^-}
 \ee
which gives, for $\bar B^0\to \pi^-\pi^+) $, $\delta_{PT} =
-36.5^{\circ} $  close to the  value
$-41.3^{\circ}$ in a  more precise determination \cite{Kou}. Similar  
determination of the strong phase could be done for $B\to K\pi$
decays by using the quantity 
$D=2(\Gamma_{\rm av}(\bar B^0\to K^-\pi^+)- \Gamma_{\rm av}(B^-\to \bar
K^0\pi^-)- T^{2})$ which is given by:
 \be
D = -4\,PT\cos(\delta_{PT})\,\cos(\gamma)
\label{D}
 \ee
The ratio $R_{K^-\pi^+}=\Gamma_{\rm as}(\bar B^0\to K^-\pi^+)/D$ is
then:
 \be
R_{K^-\pi^+}= -\tan(\delta_{PT})\tan(\gamma)
\label{Rc}
 \ee
from which we obtain~:
 \bea
&&\tan(\delta_{PT})= -\frac{R_{K^-\pi^+}}{\tan(\gamma)}\quad,\nonumber\\
&&\sin(\delta_{PT})= -\frac{R_{K^-\pi^+}}{\sqrt(\tan^{2}(\gamma) + R_{K^-\pi^+}^{2})}
\label{dPT1}
 \eea

 From the measured $B\to K\pi$ branching ratios and the QCDF expression
for $T^{2}$, we obtain $D=-5.35$, $R_{K^-\pi^+}= 0.71 $
 (in terms of the  branching ratios and in unit of $10^{-6}$)  which give, 
 \be
\tan(\delta_{PT})= -0.30,\quad  \delta_{PT}= -17^{\circ}.
\label{dPT}
 \ee
within an error of $20-30\%$, including a small theoretical uncertainty 
in the use of QCDF for $T^{2}$ which makes only a small contribution to $D$
relative to the main tree-penguin interference term. This value 
is smaller than the value $-36.5^{\circ} $ for $\delta_{PT} $ in 
$\bar B^0\to \pi^-\pi^+ $ mentioned above, but the small value
of the strong phase $\delta_{PT} $ we obtained here from 
$\bar B^0\to K^-\pi^+ $ could be due to the cancellation between 
the factorisable term, penguin-induced weak annihilation and FSI charmed meson
 intermediate states contribution to produce a negative CP asymmetry 
in $\bar B^0\to K^-\pi^+) $ decay\cite{Pham3}. 

  We now come to the ${\mathcal A}_{\rm CP} $ puzzle. As mentioned
 earlier, the solution of the puzzle requires a moderate $C/T$ ratio, but
with a strong phase $\delta_{CT}$ sufficiently large to keep the real
part of the color-suppressed tree contribution close to QCDF
prediction, like those computed for $B\to \pi\pi$ decays\cite{Beneke}. 
Then the large absorptive part could find an 
explanation from FSI effects as mentioned earlier. Indeed, as shown in
 the following, such a large strong phase for $C$ is required to produce
a large CP asymmetry  for $\bar B^0\to \bar K^0\pi^0 $ . Defining
 $R_{\bar K^0\pi^0}=\Gamma_{\rm as}(\bar B^0\to\bar K^0\pi^0 )/D $, we 
have:
 \bea
\kern -0.6cm&&R_{\bar K^0\pi^0}=
 -\frac{1}{2}\frac{C}{T}\Bigl(\sin(\delta_{CT})\,\tan(\gamma) +
 \cos(\delta_{CT})R_{K^-\pi^+}\nonumber \\
\kern -0.6cm&& + R_{W}\sin(\delta_{CT} + \delta_{T})
\sqrt(R_{K^-\pi^+}^{2} + \tan^{2}(\gamma))\Bigr)
\label{R0}
 \eea
where $R_{W}=P_{W}/P$ which is given  approximately 
by QCDF\cite{QCDF2}:
 \be
R_{W}= \frac{3}{2}\frac{f_{\pi} F^{BK}(0)}{f_{K}
  F^{B\pi}(0)}\frac{|a_{9}-a_{7}|}{|a_{4}+ a_{6}r_{\chi}|}\approx  0.13
\label{RW} 
\ee

The CP asymmetry for ${\bar B}^{0}\to\bar K^0\pi^0 $ is then
 \be
{\mathcal A}_{\rm CP}({\bar B}^{0}\to\bar K^0\pi^0)= \frac{D\,R_{\bar K^0\pi^0}}{2\,{\mathcal B}({\bar B}^{0}\to {\bar  K}^{0}\pi^{0})}
\label{ACP0} 
\ee
with  $D $  given in terms
of the  CP-averaged $B \to K\pi$ branching ratios, experimentally, $D=-5.35$, 
as mentioned above (in unit of $10^{-6}$).

A nice feature of the above expression for $R_{\bar K^0\pi^0}$   is
that it gives the CP asymmetry for $\bar B^0\to \bar K^0\pi^0 $ in terms 
of the strong phase $\delta_{CT} $, the measured $\bar B^0\to K^-\pi^+$
CP asymmetry and the weak phase $\gamma$ . For a large strong phase
$\delta_{CT} $, the $\cos(\delta_{CT})R_{K^-\pi^+} $ term is suppressed
so that the dependence of $R_{\bar K^0\pi^0}$ on $R_{K^-\pi^+} $ is weak.
There is also some dependence on $\delta_{T} $ in the electroweak
contribution to $R_{\bar K^0\pi^0}$ which could produce
a small uncertainty  on the CP asymmetry, about $10-15\%$, roughly  the size
of the electroweak penguin contribution. Thus the  
CP asymmetry for $\bar B^0\to \bar K^0\pi^0 $ depends essentially on the
strong phase of the color-suppressed tree 
contribution $\delta_{CT} $.

Numerically,  from the 
measured ${\mathcal B}(\bar B^0\to K^-\pi^+)$, the CP asymmetry 
${\mathcal A}_{\rm CP}(\bar B^0\to K^-\pi^+)$, $\gamma =67^{\circ}$, 
and taking $\delta_{T} =30^{\circ}$, we obtain:
 \be
\kern -0.3cm{\mathcal A}_{\rm CP}({\bar B}^{0}\to\bar K^0\pi^0)\kern -0.1cm=\kern -0.1cm 0.27\frac{C}{T}\bigl[1.31\sin(\delta_{CT})\kern -0.1cm + 0.44\cos(\delta_{CT})\bigr]
\label{ACP1} 
\ee

 Thus a large negative value for $\delta_{CT} $ could produce 
 a  large negative ${\mathcal A}_{\rm CP}({\bar B}^{0}\to\bar K^0\pi^0)$ 
which is needed to accommodate  the measured positive asymmetry for 
$( B^-\to K^-\pi^0)$\cite{HFAG}. For example, with 
 $\delta_{CT} =-72^{\circ}$, one would get 
 ${\mathcal A}_{\rm CP}({\bar B}^{0}\to\bar K^0\pi^0)= -0.30(C/T)$  
which implies $C/T=1/2$ 
for ${\mathcal A}_{\rm CP}({\bar B}^{0}\to\bar K^0\pi^0)=-0.15 $.
If we neglect the electroweak penguin $P_{W}$ term, we would have:
 \be
\kern -0.3cm{\mathcal A}_{\rm CP}({\bar B}^{0}\to\bar K^0\pi^0)\kern -0.1cm=\kern -0.1cm 0.27\frac{C}{T}\bigl[1.17\sin(\delta_{CT})\kern -0.1cm + 0.35\cos(\delta_{CT})\bigr]
\label{ACP01} 
\ee
independent of $\delta_{T} $. The  same value for the CP asymmetry
would implies $\delta_{CT} =-75^{\circ}$, close to the value obtained
with electroweak penguin. Thus the determination of $C/T$ 
will not be greatly affected by the electroweak penguin contribution. 
In general from QCDF one expects a small $\delta_{T}$ ,  in our
calculation  we will put $\delta_{T}=30^{\circ}$. In terms of the 
measured ${\mathcal A}_{\rm CP}({\bar B}^{0}\to\bar K^0\pi^0) $, 
from Eq.(\ref{ACP0}), $C/T$ is then:
 \be
\kern -0.3cm (\frac{C}{T})= \frac{{\mathcal A}_{\rm CP}({\bar B}^{0}\to\bar K^0\pi^0)}{
0.27\,[1.31\sin(\delta_{CT}) + 0.44\cos(\delta_{CT})]}
\label{CT} 
\ee

\begin{figure*}[t]
\begin{center}
\includegraphics[width=7cm]{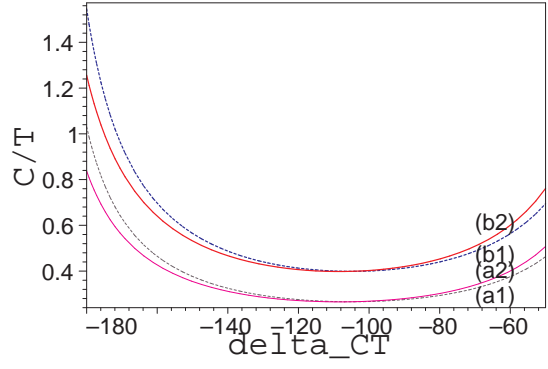}\hspace*{1cm}
\caption{The ratio $C/T$ plotted against $\delta_{CT}$, the strong
phase of the color-suppressed tree contribution $C$ . (a1,a2) are the curves
for ${\mathcal A}_{\rm CP}({\bar B}^{0}\to\bar K^0\pi^0)=-0.10 $ with $\delta_{T}$
taken to be $0.0$ and $30^{\circ}$, respectively, (b1,b2) are similar curves
for ${\mathcal A}_{\rm CP}({\bar B}^{0}\to\bar K^0\pi^0)=-0.15 $.
}
\label{fig:1cp}
\end{center}
\end{figure*}

  As shown in Fig.\ref{fig:1cp}, for  the strong phase in the range
$-(50^{\circ}-70^{\circ})$ , $C/T$ is of the order $0.3-0.4$
for an asymmetry of $-0.10$, with a larger  asymmetry  of $-0.15$ ,  
$C/T$ become larger, of the order $0.5-0.6$.
Thus  in an essentially model-independent calculation, we have shown that
a large and negative ${\bar B}^{0}\to {\bar K}^{0}\pi^{0}$ CP asymmetry, 
which is required to produce a sizable positive CP asymmetries in 
$B^{-}\to K^{-}\pi^{0}$,  implies a large 
color-suppressed tree $C$ term and its strong phase in $B\to K\pi$ decay, 
with a ratio $C/T$ of the order $0.4-0.6$ and the strong phase $\delta_{CT}$
in the range $-(50-70)^{\circ}$. Indeed, a recent analysis in QCDF
shows that the $A_{\rm CP}$ $B\to K\pi$ puzzle could be solved with
a color-suppressed tree $a_{2}$ term  large and having a large negative 
strong phase\cite{Cheng2}.
In the next section, we  will show that,  
by adding to the QCDF amplitude, an additional color-suppressed 
tree contribution with this size 
 to reverse the sign of the $B^{-}\to K^{-}\pi^{0}$
asymmetry, together with the additional  penguin 
terms (charming penguin etc.)\cite{Pham3}, indeed
good agreement with experiment is obtained  for all the  
$B\to K\pi$ branching ratios and CP asymmetries.

\section{$B\to K\pi$ DECAYS IN QCDF WITH ADDITIONAL PENGUIN AND COLOR-SUPPRESSED CONTRIBUTIONS }
 In a previous paper\cite{Pham3}, we have shown that the $B\to K\pi$
branching ratios and the ${\bar B}^{0}\to K^{+}\pi^{-}$ CP asymmetry 
could be described by QCDF with a mainly absorptive additional penguin 
terms (charming penguin etc.), with a  strength $30\%$ of the penguin term. 
However the predicted  CP asymmetry for $B^{-}\to K^{-}\pi^{0}$ is 
of the same sign and magnitude to that for ${\bar B}^{0}\to K^{-}\pi^{+}$,
in disagreement with the measured value. Therefore, to reverse the sign of
the predicted asymmetry, we need a large negative ${\bar B}^{0}\to {\bar
  K}^{0}\pi^{0}$ asymmetry and hence a color-suppressed tree term with 
large magnitude and large negative strong phase. By adding this term 
to the QCDF $B\to K\pi$ amplitudes given in our previous work\cite{Pham3}
one would obtain  correct predictions for $B\to K\pi$  branching ratios 
and CP asymmetries as will be shown below.

\begin{figure*}[t]
\begin{center}
\includegraphics[width=7cm]{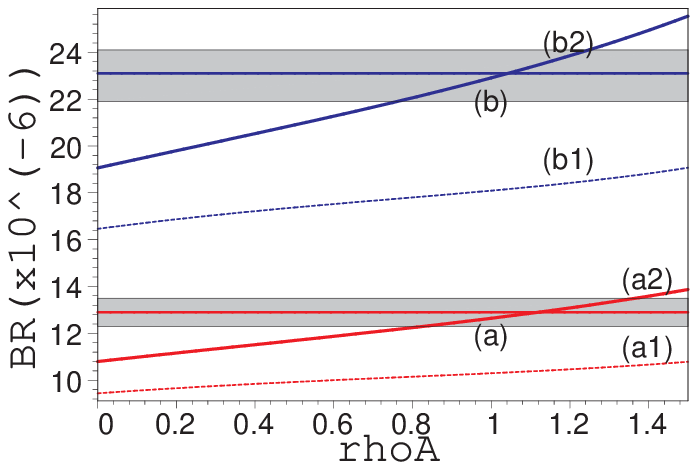}\hspace*{1cm}
\includegraphics[width=7cm]{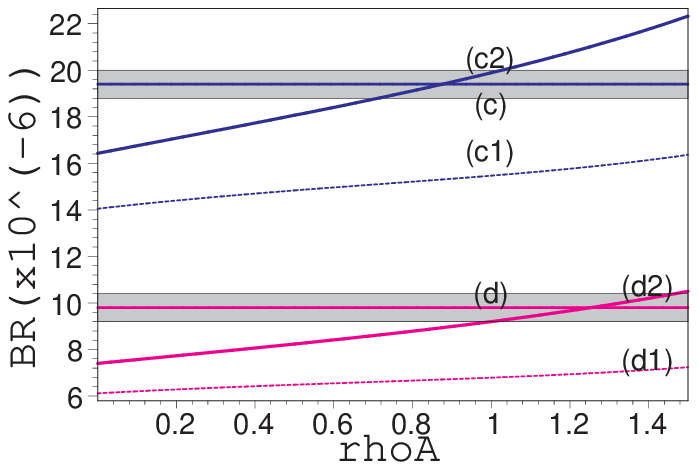} \vspace*{.5cm}
\caption{The computed and measured CP-averaged branching ratios. The
horizontal line are the measured values \cite{HFAG} with the
gray areas represent the experimental errors. (a), (b), (c), (d) in 
the left and right figure represent the values for $B^{-}\to K^{-}\pi^{0}$, 
$B^{-}\to {\bar K}^{0}\pi^{-}$, ${\bar B}^{0}\to K^{-}\pi^{+}$ and 
${\bar B}^{0}\to {\bar K}^{0}\pi^{0}$ respectively. The curves (a1)-(d1) and
(a2)-(d2) are the corresponding QCDF predicted values for
$\phi_{A}=-55^{\circ}$, without and with additional penguin-like  $\delta P$ and color-suppressed $\delta C$ contribution respectively.
}
\label{fig:1}
\end{center}
\end{figure*}

\begin{figure*}[t]
\begin{center}
\includegraphics[width=7cm]{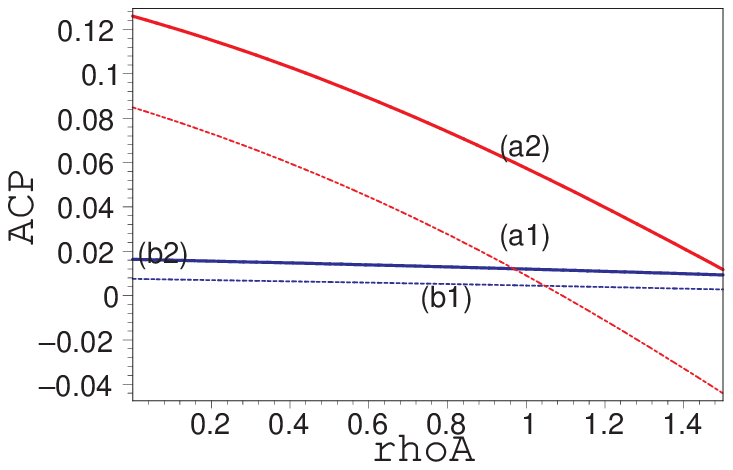}\hspace*{1cm}
\includegraphics[width=7cm]{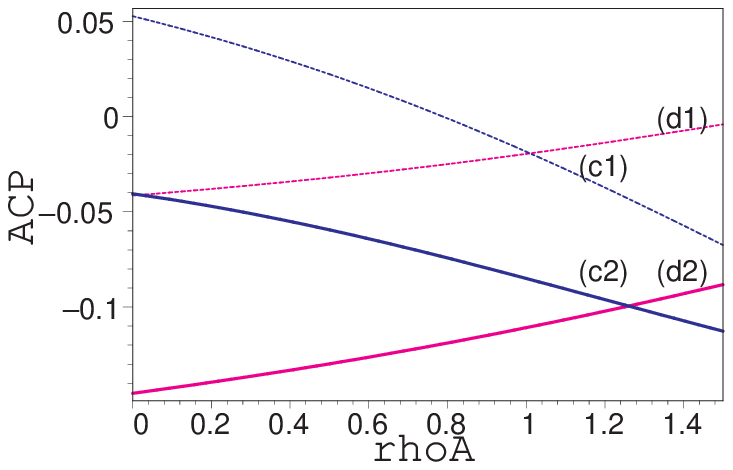}\vspace*{.5cm}
\caption{The same as in Fig.\ref{fig:1} but for the computed CP asymmetries.}
\label{fig:2}
\end{center}
\end{figure*}

 With the same hadronic, CKM parameters and the additional penguin term
$\delta P$ given in \cite{Pham3}, and writing  the color-suppressed additional 
term as $\delta C =  ra_{2}(k_{1} +i\,k_{2}) $ where $ra_{2} $ 
is  the real part of $a_{2}$, and taking $k_{1}=0$,  $k_{2}=-1.7$, 
the computed branching ratios and direct CP asymmetries, with
$\rho_{H}=1, \phi_{H}=0$ and $\phi_{A}=-55^{\circ}$ as in scenario S4 
of \cite{QCDF2} are shown in Fig.\ref{fig:1} and Fig.\ref{fig:2} 
as function of $\rho_{A}$. For convenience we  also give in 
Table \ref{tab-res1} and Table \ref{tab-res2} the computed values 
at $\rho_{A}=1$ as in S4 with and without the additional 
penguin-like $\delta P$ and color-suppressed 
$\delta C$ contribution.  We see that with these additional 
contributions, all the branching ratios and CP asymmetries are in good
agreement with the measured values. In particular, the 
${\bar B}^{0}\to {\bar K}^{0}\pi^{0}$ branching ratio is slightly larger than 
the previous predicted value of $8.9\times 10^{-6} $ due to the 
additional $\delta C$ contribution and is  closer to  experiment, while 
other predicted branching ratios remain practically unchanged. 

\begin{center}
\begin{table}[h]
\begin{tabular}{|c|c|c|c|}

\hline
 Decay & $\delta P=0$&$\delta P\not \!=0$   & Exp \cite{HFAG} \\
Modes & $\delta C=0$&$\delta C\not \!=0$   &  \\
\hline
$B^{-}\to \pi^{-}\pi^{0}$ &$5.7$ &$5.7$ & $5.59\pm 0.4$\\
$B^{-}\to K^{-}\pi^{0}$ &$10.3$ &$12.6$ & $12.9\pm 0.6$\\
$B^{-}\to {\bar K}^{0}\pi^{-}$&$18.1$ &$22.9$ & $23.1\pm 1.0$\\
$\bar{B}^{0}\to  K^{-}\pi^{+}$ &$15.5$ &$19.9$ & $19.4 \pm 0.6$\\
${\bar B}^{0}\to {\bar K}^{0}\pi^{0}$&$6.8$ &$9.2$ & $9.8\pm 0.6$\\
\hline

\end{tabular}
\caption{ The CP-averaged $B\to K\pi$
Branching ratios in unit of $ 10^{-6}$  in QCDF with and without
additional penguin-like  $\delta P$ and color-suppressed 
$\delta C$ contribution and with
 $\rho_{A}=1.0$, $\phi_{A}= -55^{\circ}$}\label{tab-res1}
\end{table}
\end{center}

\begin{center}
\begin{table}[h]
\begin{tabular}{|c|c|c|c|}
\hline
Decay & $\delta P=0$&$\delta P\not \!=0$   & Exp \cite{HFAG} \\
Modes & $\delta C=0$&$\delta C\not \!=0$   &  \\
\hline
$B^{-}\to \pi^{-}\pi^{0}$ &$0.0$ &$0.0$ & $0.06\pm 0.05$\\
$B^{-}\to K^{-}\pi^{0}$ &$0.01$ &$0.06$ & $0.05\pm 0.025$\\
$B^{-}\to {\bar K}^{0}\pi^{-}$&$0.004$ &$0.01$ & $-0.009\pm 0.025$\\
$\bar{B}^{0}\to  K^{-}\pi^{+}$ &$-0.02$ &$-0.08$ & $-0.098^{+0.012}_{-0.010}$\\
${\bar B}^{0}\to {\bar K}^{0}\pi^{0}$&$-0.02$ &$-0.11$ & $-0.01\pm 0.10$\\
\hline
\end{tabular}
\caption{ The direct $B\to K\pi$ CP asymmetries  in QCDF with and without
additional penguin-like contribution $\delta P$ and color-suppressed 
$\delta C$ contribution and with
 $\rho_{A}=1.0$, $\phi_{A}= -55^{\circ}$}\label{tab-res2}
\end{table}
\end{center}
In our previous work\cite{Pham3}, we give predictions for 
the ${\bar B}^{0}\to {\bar
  K}^{0}\pi^{0} $, $B^{-}\to   K^{-}\pi^{0} $ and ${\bar B}^{0}\to
  K^{-}\pi^{+} $ branching ratios in terms of the computed 
differences $2{\mathcal B}({\bar B}^{0}\to {\bar K}^{0}\pi^{0})-
r_{b}{\mathcal B}(B^{-}\to {\bar K}^{0}\pi^{-}) $, 
$2\,r_{b}{\mathcal B}(B^{-}\to
K^{-}\pi^{0})- {\mathcal B}({\bar B}^{0}\to  K^{-}\pi^{+})$, 
${\mathcal B}({\bar B}^{0}\to K^{-}\pi^{+} )-
r_{b}{\mathcal B}(B^{-}\to {\bar K}^{0}\pi^{-}) $ and the measured 
${\bar B}^{0}\to K^{-}\pi^{+} $ and $B^{-}\to {\bar K}^{0}\pi^{-} $
branching ratios. The good agreement with experiment shows that 
QCDF could describe rather well the electroweak penguin  contribution.
We  give here similar predictions with the additional  
color-suppressed term included (in unit of $10^{-6}$):
\bea
\kern -0.6 cm&&{\mathcal B}({\bar B}^{0}\to {\bar K}^{0}\pi^{0})= 9.3\pm 0.3, 
\label{BK0}\\
\kern -0.6 cm&&{\mathcal B}(B^{-}\to   K^{-}\pi^{0})= 12.4\pm 0.3.
\label{BK1}\\
\kern -0.6 cm&&{\mathcal B}({\bar B}^{0}\to  K^{-}\pi^{+})= 20.1\pm 0.6, 
\label{BK3}
\eea
We see that because of the large color-suppressed contribution, 
the ${\bar B}^{0}\to {\bar K}^{0}\pi^{0} $ predicted
branching ratio is larger than the previous predicted value $(9.0\pm 0.3)\times
10^{-6}$ and is closer to experiment, the other two predicted 
branching ratios are almost unchanged and are in good agreement with
experiment within the current accuracy.
\bigskip
\section{Conclusion}
 By adding mainly absorptive 
additional  penguin-like  and  color-suppressed tree terms  to the QCDF
$B\to K\pi$ decay amplitudes, we show that QCDF could 
successfully predict the $B\to K\pi$ branching ratios and  
CP asymmetries. In particular, with a large negative strong phase for
the color-suppressed tree contribution, we obtain the 
correct magnitude
and sign for the ${\bar B}^{0}\to K^{-}\pi^{+} $ and 
 $B^{-}\to K^{-}\pi^{0}$ CP asymmetry, and a large negative asymmetry for
${\bar B}^{0}\to {\bar K}^{0}\pi^{0} $. Confirmation of these CP asymmetries  
by new measurements and the measurement of  
${\bar B}^{0}\to \pi^{0}\pi^{0} $ CP asymmetry\cite{Kou} would be 
an evidence for a large $C/T$ ratio and a large 
strong phase in $B\to \pi\pi$ and $B\to K\pi$ decays.
\bigskip
\begin{center}
{\bf Acknowledgments} \end{center}
I would like to thank  Hai-Yang Cheng for useful discussions. This work was 
supported in part by the EU contract No. MRTN-CT-2006-035482, "FLAVIAnet".


\end{document}